# Circular multi-party quantum private comparison with $n$-level single-particle states

Chong-Qiang Ye, Tian-Yu Ye*

College of Information & Electronic Engineering, Zhejiang Gongshang University, Hangzhou 310018, P.R.China
*E-mail：happyyty@aliyun.com

**Abstract:** In this paper, a novel multi-party quantum private comparison (MQPC) protocol for equality comparison with $n$-level single-particle states is constructed, where the encoded particles are transmitted in a circular way. Here, $n$ parties employ the qudit shifting operation to encode their private secrets and can compare the equality of their private secrets within one time execution of protocol. The proposed MQPC protocol can overcome both the outside attack and the participant attack. Specially, each party's secret can be kept unknown to other parties and the third party (TP).

## 1 Introduction

Classical cryptography, whose security relies on the computation complexity of mathematical problems, is vulnerable to the powerful computing abilities of quantum parallel computation. Fortunately, quantum cryptography, invented by Bennett and Brassard [1] in 1984, can gain its unconditional security through the laws of quantum mechanics. It has attracted lots of attention and established many interesting branches, such as QKD [1-7], quantum secure direct communication (QSDC) [8-11], quantum secret sharing (QSS) [12-18] *etc*.

Private comparison protocol can be traced to the millionaires' problem proposed by Yao [19]. In the millionaires' problem, two millionaires want to judge who is richer without knowing each other's actual property. Afterward, Boudot *et al.* [20] suggested a private comparison protocol to determine whether two millionaires are equally rich. Unfortunately, Lo [21] pointed out that it is impossible to construct a secure equality function in a two-party scenario. Hence, some additional assumptions, such as a third party (TP), should be taken into account for private comparison.

Quantum private comparison (QPC), which can be seen as the generalization of classical private comparison into the realm of quantum mechanics, was first proposed by Yang and Wen [22]. The goal of QPC is to decide whether the private inputs from different parties are equal or not by utilizing the principles of quantum mechanics on the basis that none of their genuine contents are leaked out. In recent years, the design and analysis of QPC protocols have successfully attracted much attention. As a result, many two-party QPC protocols [22-35] have been proposed.

However, within one time execution of protocol, each of the above two-party QPC protocols can only accomplish the equality comparison of secrets from two parties. If the two-party QPC protocol is used to solve the multi-party equality comparison problem, it will have to be executed with $(n-1) \sim n(n-1)/2$ times, which will result in the reduction of efficiency. Here, $n$ is the number of parties. Fortunately, in 2013, by utilizing multi-particle GHZ class states, Chang *et al.*[36] proposed the multi-party quantum private comparison (MQPC) protocol for the first time to achieve arbitrary pair's comparison of equality among $n$ parties within one time execution of protocol. Afterward, many MQPC protocols within the multi-level system [37-43] have been quickly constructed to accomplish the equality comparison among $n$ parties within one time execution of protocol. For example, in 2014, Liu *et al.* [37] proposed a MQPC protocol using $d$ dimensional basis states without entanglement swapping; and in 2017, Ji and Ye [39] presented a MQPC protocol based on the entanglement swapping of $d$-level cat states and $d$-level Bell states.

Based on the above analysis, in this paper, we propose a novel MQPC protocol by using $d$-level single-particle states, where the encoded particles are transmitted in a circular way. The proposed protocol can compare the equality of secrets from $n$ parties with the help of a semi-honest TP within one time execution of protocol. Here, TP is semi-honest in the sense that she may misbehave on her own but is not allowed to collude with anyone else.

## 2  Preliminary knowledge

In a $n$-level quantum system, a common basis of single photons can be expressed as

$$T_1 = \{|k\rangle\}, \quad k \in \{0,1,\cdots,n-1\}. \tag{1}$$

Each element in the set $T_1$ is orthogonal to the others. Performing the $n$ th order discrete quantum Fourier transform $F$ on each state in $T_1$ can form another basis which can be described as Eq.(2):

$$T_2 = \{F|k\rangle\} = \left\{ \frac{1}{\sqrt{n}} \sum_{j=0}^{n-1} \omega^{jk} |j\rangle \right\}. \tag{2}$$

Here, $\omega = e^{\frac{2\pi i}{n}}$ and $k \in \{0,1,\cdots,n-1\}$. Each element in the set $T_2$ is also orthogonal to the others. Apparently, the two sets, $T_1$ and $T_2$, are two common conjugate bases. The unitary operation shown in Eq.(3), represents the qudit shifting operation, where the symbol "$\oplus$" denotes the addition modulo $n$ and $m \in \{0,1,\ldots,n-1\}$. It is easy to verify that after the particle $|k\rangle$ is performed with the qudit shifting operation $U_m$, its state is converted into $|k \oplus m\rangle$.

$$U_m = \sum_{r=0}^{n-1} |r \oplus m\rangle\langle r|. \tag{3}$$

**Theorem 1.** *After the particle $F|k\rangle$ is performed with the qudit shifting operation $U_m$, its state is kept unchanged.*

**Proof.** (i) After the particle $F|k\rangle$ is performed with the qudit shifting operation $U_1$, its state evolves into

$$U_1 F|k\rangle = \left( \sum_{r=0}^{n-1} |r \oplus 1\rangle\langle r| \right) \frac{1}{\sqrt{n}} \sum_{j=0}^{n-1} \omega^{jk} |j\rangle$$

$$= \left( \sum_{r=0}^{n-2} |r+1\rangle\langle r| + |0\rangle\langle n-1| \right) \frac{1}{\sqrt{n}} \sum_{j=0}^{n-1} \omega^{jk} |j\rangle$$

$$= \left( \sum_{r=0}^{n-2} |r+1\rangle\langle r| \right) \frac{1}{\sqrt{n}} \sum_{j=0}^{n-1} \omega^{jk} |j\rangle + \frac{1}{\sqrt{n}} \omega^{(n-1)k} |0\rangle$$

$$= \frac{1}{\sqrt{n}} \sum_{r=0}^{n-2} \omega^{rk} |r+1\rangle + \frac{1}{\sqrt{n}} \omega^{(n-1)k} |0\rangle. \tag{4}$$

By replacing $r+1$ with $t$ $(t \in \{1,2,\ldots,n-1\})$, Eq.(4) will evolve into

$$U_1 F|k\rangle = \frac{1}{\sqrt{n}} \sum_{t=1}^{n-1} \omega^{(t-1)k} |t\rangle + \frac{1}{\sqrt{n}} \omega^{(n-1)k} |0\rangle$$

$$= \omega^{-k} \frac{1}{\sqrt{n}} \sum_{t=1}^{n-1} \omega^{tk} |t\rangle + \frac{1}{\sqrt{n}} \omega^{(n-1)k} |0\rangle$$

$$= \omega^{-k} \left( \frac{1}{\sqrt{n}} \sum_{t=1}^{n-1} \omega^{tk} |t\rangle + \frac{1}{\sqrt{n}} \omega^{nk} |0\rangle \right)$$

$$= \omega^{-k} \left( \frac{1}{\sqrt{n}} \sum_{t=0}^{n-1} \omega^{tk} |t\rangle \right)$$

$$= \omega^{-k} F|k\rangle. \tag{5}$$

(ii) It is apparent that performing the qudit shifting operation $U_m$ is equivalent to performing the qudit shifting operation $U_1$ $m$ times. Thus, it can be easily obtained from Eq.(5) that

$$U_m F|k\rangle = \omega^{-mk} F|k\rangle. \tag{6}$$

It can be concluded now that after the global factor $\omega^{-mk}$ is ignored, the qudit shifting operation $U_m$ can not change the state of $F|k\rangle$. ∎

## 3 The proposed MQPC protocol

Now, let us give an explicit description for the proposed MQPC protocol. Suppose that there are $n$ parties, $P_1, P_2, \ldots, P_n$, where $P_i$ $(i=1,2,\ldots,n)$ has the secret $x_i$. The binary representation of $x_i$ is $(x_i^1, x_i^2, \ldots, x_i^L)$, namely, $x_i = \sum_{l=1}^{L} x_i^l 2^{l-1}$, where $x_i^l \in \{0,1\}$ for $l = 1,2,\ldots,L$. They want to determine whether all of their secrets are equal or not with the help of the semi-honest TP. TP is semi-honest in the sense that she may misbehave on her own but is not allowed to conspire with anyone else. Note that this kind of semi-honest TP belongs to the second semi-honest model of TP in Ref.[25]. TP and $P_i$ $(i=1,2,\ldots,n)$ share a key sequence $Q_i$ of length $L$ beforehand, through a secure QKD protocol, where $Q_i = (q_i^1, q_i^2, \ldots, q_i^L)$, $q_i^l \in \{0,1,\ldots,n-1\}$ and $l = 1,2,\ldots,L$. For description convenience, we use $P_0$ to represent TP hereafter.

All participants execute the following steps together to accomplish the private comparison task.

**Step 1:** $P_0$ prepares $L$ single particles, $|k_1\rangle, |k_2\rangle, \ldots, |k_L\rangle$, which are randomly chosen from the set $T_1$, and uses these particles to construct the particle sequence $H$. In the meanwhile, $P_0$ prepares another $L$ single particles, $F|k_1\rangle, F|k_2\rangle, \ldots, F|k_L\rangle$, which are randomly chosen from the set $T_2$, and uses these particles to construct the particle sequence $V$. Here, $k_l \in \{0,1,\ldots,n-1\}$ and $l = 1,2,\ldots,L$. Then, $P_0$ takes the particle $|k_l\rangle$ out from sequence $H$ and the particle $F|k_l\rangle$ out from sequence $V$ to construct the particle group $(|k_l\rangle, F|k_l\rangle)$, which is denoted as $G_l^0$ for convenience. It should be noted that in each $G_l^0$, the positions of particles $|k_l\rangle$ and $F|k_l\rangle$ are random which are only known by $P_0$. Afterward, $P_0$ prepares $2L$ decoy photons, which are randomly chosen from the set $T_1$ or $T_2$, and inserts them into $G_1^0, G_2^0, \ldots, G_L^0$ at random positions to construct the new sequence $S_0$. Finally, $P_0$ sends $S_0$ to $P_1$.

**Step 2:** For $i = 1,2,\ldots,n$:

After confirming that $P_i$ has received all particles from $P_{i-1}$, $P_{i-1}$ checks the security of the transmission of $S_{i-1}$ with $P_i$. Firstly, $P_{i-1}$ announces the positions and the preparation bases of decoy photons which are chosen from the set $T_2$. According to the announced information, $P_i$ uses the base $T_2$ to measure the corresponding decoy photons and returns the measurement results to $P_{i-1}$. By comparing the measurement results of decoy photons with their corresponding initial prepared states, $P_{i-1}$ can calculate the error rate. If the error rate is greater than the threshold value, the protocol will be terminated and restarted from Step 1. Otherwise, $P_{i-1}$ tells $P_i$ the positions and the preparation bases of decoy photons which are chosen from the set $T_1$. Then $P_i$ uses the base $T_1$ to measure the corresponding decoy photons and returns the measurement results to $P_{i-1}$. By comparing the measurement results of decoy photons with their corresponding initial prepared states, $P_{i-1}$ can calculate the error rate. If the error rate is greater than the threshold value, the protocol will be terminated and restarted from Step 1; otherwise, the protocol will be proceeded.

After removing the decoy photons in $S_{i-1}$, in order to encode her secret binary bit $x_i^l$, $P_i$ performs the qudit shifting operation $U_{q_i^l \oplus x_i^l} = \sum_{r=0}^{n-1} |r \oplus q_i^l \oplus x_i^l\rangle\langle r|$ on the two particles $|k_l \oplus q_1^l \oplus q_2^l \oplus \ldots \oplus q_{i-1}^l \oplus x_1^l \oplus x_2^l \oplus \ldots \oplus x_{i-1}^l\rangle$ and $F|k_l\rangle$ in $G_l^{i-1}$. Note that only $P_i$ and $P_0$ know the genuine value of $q_i^l$. Here, $q_i^l \in \{0,1,\ldots,n-1\}$ and $l = 1,2,\ldots,L$. As a result, the state of $|k_l \oplus q_1^l \oplus q_2^l \oplus \ldots \oplus q_{i-1}^l \oplus x_1^l \oplus x_2^l \oplus \ldots \oplus x_{i-1}^l\rangle$ is changed into $|k_l \oplus q_1^l \oplus q_2^l \oplus \ldots \oplus q_i^l \oplus x_1^l \oplus x_2^l \oplus \ldots \oplus x_i^l\rangle$, but the state of $F|k_l\rangle$ is kept unchanged according to Theorem 1. The sequence $G_1^{i-1}, G_2^{i-1}, \ldots, G_L^{i-1}$ after the encoding operations of $P_i$ is denoted as $G_1^i, G_2^i, \ldots, G_L^i$. Then, $P_i$ prepares $2L$ decoy photons, which are randomly chosen from the set $T_1$ or $T_2$, and inserts them into $G_1^i, G_2^i, \ldots, G_L^i$ at random positions to construct a new sequence $S_i$. Finally, $P_i$ sends $S_i$ to $P_{i+1}$. It should be pointed out that $P_n$ sends $S_n$ to $P_0$.

**Step 3:** After confirming that $P_0$ has received all particles from $P_n$, $P_n$ checks the security of the transmission of $S_n$ with $P_0$. Firstly, $P_n$ announces the positions and the preparation bases of decoy photons which are chosen from the set $T_2$. According to the announced information, $P_0$ uses the base $T_2$ to measure the corresponding decoy photons and returns the measurement results to $P_n$. By comparing the measurement results of decoy photons with their corresponding initial prepared states, $P_n$ can calculate the error rate. If the error rate is greater than the threshold value, the protocol will be terminated and restarted from Step 1. Otherwise, $P_n$ tells $P_0$ the positions and the preparation bases of decoy photons which are chosen from the set $T_1$. Then $P_0$ uses the base $T_1$ to measure the corresponding decoy photons and returns the measurement results to $P_n$. By comparing the measurement results of decoy photons with their corresponding initial prepared states, $P_n$ can calculate the error rate. If the error rate is greater than the threshold value, the protocol will be terminated and restarted from Step 1; otherwise, the protocol will be proceeded.

$P_0$ removes the decoy photons in $S_n$. $P_0$ can know the correct measuring bases of the remaining particles, as $P_i$ ($i = 1,2,\ldots,n$) did not change the orders of the initial particles prepared by $P_0$ in Step 1 when performing the qudit shifting operation to encode her secret. Then $P_0$ uses the base $T_2$ to measure the particles from the set $T_2$ and compares the measurement results with their corresponding initial prepared states. If the error rate is greater than the threshold value, the protocol will be terminated and restarted from Step 1; otherwise, the protocol will be proceeded to the next step.

**Step 4:** $P_0$ uses the base $T_1$ to measure the particle $|k_l \oplus q_1^l \oplus q_2^l \oplus \ldots \oplus q_n^l \oplus x_1^l \oplus x_2^l \oplus \ldots \oplus x_n^l\rangle$ and obtains the value of

$k_l \oplus q_1^l \oplus q_2^l \oplus \ldots \oplus q_n^l \oplus x_1^l \oplus x_2^l \oplus \ldots \oplus x_n^l$. Note that $P_0$ knows the values of $q_1^l, q_2^l, \ldots, q_n^l$. Here, $l = 1, 2, \ldots, L$. Then $P_0$ calculates

$$R_l = q_1^l \oplus q_2^l \oplus \cdots \oplus q_n^l, \qquad (7)$$

$$\begin{aligned} D_l &= [k_l \oplus q_1^l \oplus q_2^l \oplus \ldots \oplus q_n^l \oplus x_1^l \oplus x_2^l \oplus \ldots \oplus x_n^l \oplus (n - k_l) - R_l] \bmod n \\ &= (x_1^l + x_2^l + \ldots + x_n^l) \bmod n \\ &= x_1^l \oplus x_2^l \oplus \ldots \oplus x_n^l. \end{aligned} \qquad (8)$$

If $D_l = 0$ for $l = 1, 2, \ldots, L$, $P_0$ will conclude that secrets of $P_1, P_2, \ldots, P_n$ are same; otherwise, she will conclude that their secrets are not same. Finally, $P_0$ announces the comparison result to $P_1, P_2, \ldots, P_n$ through the classical channel.

## 4  Analysis

In this section, we analyze the output correctness and the security of the proposed MQPC protocol. Firstly, we show that the output of the proposed MQPC protocol is correct. Secondly, we show that both the outside attack and the participant attack are invalid to the proposed MQPC protocol.

### 4.1  Output correctness

There are $n$ parties named $P_1, P_2, \ldots, P_n$, where $P_i$ $(i = 1, 2, \ldots, n)$ has a secret $x_i$. The binary representation of $x_i$ is $(x_i^1, x_i^2, \ldots, x_i^L)$, namely, $x_i = \sum_{l=1}^{L} x_i^l 2^{l-1}$, where $x_i^l \in \{0, 1\}$ for $l = 1, 2, \ldots, L$. We take the $l^{\text{th}}$ bit of $x_i$, i.e., $x_i^l$, for example to illustrate the output correctness. For $i = 1, 2, \ldots, n$, $P_i$ encodes $x_i^l$ by performing the qudit shifting operation $U_{q_i^l \oplus x_i^l} = \sum_{r=0}^{n-1} |r \oplus q_i^l \oplus x_i^l\rangle\langle r|$ on the particle $|k_l \oplus q_1^l \oplus q_2^l \oplus \ldots \oplus q_{i-1}^l \oplus x_1^l \oplus x_2^l \oplus \ldots \oplus x_{i-1}^l\rangle$. After all parties finish the encoding operations on the particle $|k_l\rangle$, its final state becomes $|k_l \oplus q_1^l \oplus q_2^l \oplus \ldots \oplus q_n^l \oplus x_1^l \oplus x_2^l \oplus \ldots \oplus x_n^l\rangle$. Then, $P_0$ uses the base $T_1$ to measure it and gets the value of $k_l \oplus q_1^l \oplus q_2^l \oplus \ldots \oplus q_n^l \oplus x_1^l \oplus x_2^l \oplus \ldots \oplus x_n^l$. Afterward, $P_0$ calculates $D_l$. According to Eq.(8), apparently, if and only if $x_1^l = x_2^l = \ldots = x_n^l$, then $D_l = 0$; otherwise, $D_l \neq 0$. Finally, $P_0$ publicly announces the comparison result of equality to $P_1, P_2, \ldots, P_n$.

Based on the above analysis, it is easy to obtain that if and only if $D_1 = D_2 = \ldots = D_L = 0$, we have $x_1 = x_2 = \ldots = x_n$; otherwise, $x_1, x_2, \ldots, x_n$ are not same. It can be concluded that the output of the proposed MQPC protocol is correct.

### 4.2  Security

#### 4.2.1  The outside attack

Here, we analyze the possibility for an outside eavesdropper to steal the secrets of $n$ parties.

In the proposed MQPC protocol, there are the qudit transmissions via quantum channels in Steps 1 and 2. An outside eavesdropper may utilize these qudit transmissions to obtain something useful about the secrets of $n$ parties through launching some famous attacks such as the intercept-resend attack, the measure-resend attack and the entangle-measure attack. However, these steps use the decoy photon technique [44,45] to ensure the security of qudit transmissions. This kind of eavesdropping check method can be considered as a variant of its counterpart of the BB84 protocol [1] which has been proven to be unconditionally secure [46]. The effectiveness of decoy photon technology within 2-level quantum system against the intercept-resend attack, the measure-resend attack and the entangle-measure attack has been validated in detail in Refs.[47,48].

It is straightforward that the decoy photon technology within $n$-level quantum system is also effective against these famous attacks. Therefore, an outside eavesdropper cannot obtain something useful about the secrets without being discovered by the eavesdropping check processes in Steps 2 and 3.

In Step 4, $P_0$ announces the comparison result of equality. Even though an outside eavesdropper may hear of it, this information is still useless for her to get the secrets.

In addition, it needs to be pointed out that in order to resist the invisible photon eavesdropping Trojan horse attack [49], the receiver should insert a filter in front of her devices to filter out the photon signal with an illegitimate wavelength [50,51]. Moreover, in order to resist the delay-photon Trojan horse attack [50,52], the receiver should employ a photon number splitter (PNS:50/50) to split each sample quantum signal into two pieces and measure the signals after the PNS with proper measuring bases [50,51]. If the multiphoton rate is unreasonably high, this attack will be detected.

#### 4.2.2 The participant attack

In 2007, Gao *et al*. [53] pointed out for the first time that a dishonest participant's attack is generally more powerful and should be paid more attention to. Hereafter, this kind of attack is always called as the participant attack, and has aroused much interest in the cryptanalysis of quantum crytography [54-56]. In this subsection, we consider two cases of participant attack. On one hand, we analyze the attack from one or more dishonest parties; and on the other hand, we analyze the participant attack from semi-honest $P_0$.

**Case 1: The participant attack from one or more dishonest parties.**

This case involves two situations. One is that one dishonest party wants to extract other parties' secrets; and the other one is that two or more dishonest parties conclude together to extract other parties' secrets.

**(a) The participant attack from one dishonest party**

In the proposed MQPC protocol, $P_i$ ($i=1,2,\ldots,n$) receives particles from $P_{i-1}$ and sends particles to $P_{i+1}$. It should be pointed out that $P_n$ sends particles to $P_0$. Without loss of generality, in this situation, we suppose that $P_2$ is the only dishonest party who wants to steal other parties' secrets.

Firstly, consider the circumstance that $P_2$ wants to extract the secret of $P_1$. After the eavesdropping check processes between $P_1$ and $P_2$, in order to extract the secret of $P_1$, $P_2$ removes the decoy photons first, then measures the remaining particles $|k_l \oplus q_1^l \oplus x_1^l\rangle$ ($l=1,2,\ldots,L$) and $F|k_l\rangle$ in her hand with the base $T_1$. However, $P_2$ still cannot obtain $x_1^l$ because of the following aspects: on one hand, she has no knowledge about the genuine orders of particles $|k_l \oplus q_1^l \oplus x_1^l\rangle$ and $F|k_l\rangle$; and on the other hand, she still cannot know the values of $k_l$ and $q_1^l$. More seriously, she will inevitably be caught by $P_0$ during the final eavesdropping check process of Step 3, because her attack alters the state of particle $F|k_l\rangle$. Thus, $P_2$ cannot extract the secret of $P_1$.

Secondly, consider the circumstance that $P_2$ wants to extract the secret of $P_3$. In order to extract the secret of $P_3$, $P_2$ may take the following actions: $P_2$ prepares fake particles with the base $T_1$ and sends $P_3$ these fake ones instead of those from $P_1$; and after $P_3$ finishes encoding her secret, $P_2$ intercepts the particles from $P_3$ to $P_4$, measures them with the base $T_1$ and sends the measured particles to $P_4$. However, $P_2$ still cannot get the secret of $P_3$, as she has no knowledge about the positions of her fake particles in the particle sequence from $P_3$ to $P_4$. More seriously, her attack will be detected by the eavesdropping check processes performed by $P_3$ and $P_4$, as her attack inevitably alters the states of decoy photons from the set $T_2$ produced by $P_3$. Thus, $P_2$ cannot extract the secret of $P_3$.

Thirdly, consider the circumstance that $P_2$ wants to extract the secret of $P_j$ ($j \neq 1$ and $j \neq 3$). In the proposed MQPC protocol, there is not any particle directly transmitted between $P_2$ and $P_j$. In order to obtain the secret of $P_j$, $P_2$ may try to launch her attacks on the transmitted particles from $P_j$ to $P_{j+1}$. In this way, she essentially acts as an outside eavesdropper. As analyzed in Sect.4.2.1, she will inevitably be caught, as she has no knowledge about the positions and the preparation bases of transmitted decoy photons prepared by $P_j$. Thus, $P_2$ cannot extract the secret of $P_j$.

It can be concluded now that one dishonest party cannot obtain other parties' secrets.

**(b) The participant attack from two or more dishonest parties**

When the number of dishonest parties concluding together is $n-1$, the extreme case of the participant attack from two or more dishonest parties arise, as in this case, the union of dishonest parties becomes the most powerful.

Firstly, consider the circumstance that $P_2, P_3, \ldots, P_n$ collude together to get the secret of the honest $P_1$. After the eavesdropping check processes with $P_1$, in order to extract the secret of $P_1$, they remove the decoy photons first, then measure the remaining particles $|k_l \oplus q_1^l \oplus x_1^l\rangle$ ($l = 1, 2, \ldots, L$) and $F|k_l\rangle$ with the base $T_1$. However, they still cannot obtain $x_1^l$ because they have no knowledge about the genuine orders of particles $|k_l \oplus q_1^l \oplus x_1^l\rangle$ and $F|k_l\rangle$ and the values of $k_l$ and $q_1^l$. More seriously, they will inevitably be caught by $P_0$ during the final eavesdropping check process of Step 3, because their attack alters the state of particle $F|k_l\rangle$. Thus, $P_2, P_3, \ldots, P_n$ cannot extract the secret of $P_1$.

Secondly, consider the circumstance that $P_1, P_2, \ldots, P_{n-1}$ collude together to get the secret of the honest $P_n$. In this circumstance, $P_1, P_2, \ldots, P_{n-1}$ prepare fake particles with the base $T_1$ and send $P_n$ these fake ones instead of those from $P_0$; and after $P_n$ finishes encoding her secret, they intercept the particles from $P_n$ to $P_0$, measure them with the base $T_1$ and send the measured particles to $P_0$. However, they still cannot get the secret of $P_n$, as they have no knowledge about the positions of their fake particles in the particle sequence from $P_n$ to $P_0$. More seriously, their attack will be detected by the eavesdropping check processes performed by $P_n$ and $P_0$, as their attack inevitably alters the states of decoy photons from the set $T_2$ produced by $P_n$. Thus, $P_1, P_2, \ldots, P_{n-1}$ cannot extract the secret of $P_n$.

Thirdly, consider the circumstance that $P_1, \ldots, P_{m-1}, P_{m+1}, \ldots, P_n$ collude together to get the secret of the honest $P_m$, where $m = 2, 3, \ldots, n-1$. In this circumstance, they prepare fake particles with the base $T_1$ and send $P_m$ these fake ones instead of those from $P_0$; after $P_m$ finishes encoding her secret, $P_m$ sends the encoded particles together with her prepared decoy photons to $P_{m+1}$; after the eavesdropping check processes with $P_m$, they remove the decoy photons, use the base $T_1$ to measure the remaining particles and get the qudit shifting operations of $P_m$; hereafter, they perform the same qudit shifting operations as $P_m$ on the particles from $P_0$, and send these particles to $P_0$ together with their prepared decoy photons. In this way, they can obtain the value of $q_m^l \oplus x_m^l$ ($l = 1, 2, \ldots, L$), and this attack will not be detected by $P_0$. However, this information is still helpless for them to obtain $x_m^l$, as they have no knowledge about $q_m^l$. As a result, $P_1, \ldots, P_{m-1}, P_{m+1}, \ldots, P_n$ cannot extract the secret of $P_m$. Thus, the dishonest $P_1, \ldots, P_{m-1}, P_{m+1}, \ldots, P_n$ cannot obtain the secret of $P_m$.

It can be concluded now that the $n-1$ dishonest parties cannot obtain the secret of the left one party.

**Case 2: The participant attack from semi-honest $P_0$**

In the proposed MQPC protocol, $P_0$ may perform any possible attack except colluding with any party.

Firstly, consider the circumstance that $P_0$ wants to extract the secret of $P_1$. In order to extract the secret of $P_1$, $P_0$ may take the following actions: $P_0$ sends her normally prepared particles to $P_1$; and after $P_1$ finishes encoding her secret, $P_0$ intercepts the particles from $P_1$ to $P_2$, measures them with the base $T_1$ and sends the measured particles to $P_2$. However, $P_0$ still cannot get the secret of $P_1$, as she has no knowledge about the positions of her prepared particles in the particle sequence from $P_1$ to $P_2$. More seriously, her attack will be detected by the eavesdropping check processes performed by $P_1$ and $P_2$, as her attack inevitably alters the states of decoy photons from the set $T_2$ produced by $P_1$. Thus, $P_0$ cannot extract the secret of $P_1$.

Secondly, consider the circumstance that $P_0$ wants to extract the secret of $P_r$, where $r = 2, 3, \ldots, n-1$. In the proposed MQPC protocol, there is not any particle directly transmitted between $P_0$ and $P_r$. In order to obtain the secret of $P_r$, $P_0$ may try to launch her attacks on the transmitted particles from $P_r$ to $P_{r+1}$. In this way, she essentially acts as an outside eavesdropper. As analyzed in Sect.4.2.1, she will inevitably be caught, as she has no knowledge about the positions and the preparation bases of transmitted decoy photons prepared by $P_r$. Thus, $P_0$ cannot extract the secret of $P_r$.

Thirdly, consider the circumstance that $P_0$ wants to extract the secret of $P_n$. After the eavesdropping check processes in Step 3, in order to extract the secret of $P_n$, $P_0$ measures the remaining particle $|k_l \oplus q_1^l \oplus q_2^l \oplus \ldots \oplus q_n^l \oplus x_1^l \oplus x_2^l \oplus \ldots \oplus x_n^l\rangle$ ($l = 1, 2, \ldots, L$) in her hand with the base $T_1$ to get the value of $k_l \oplus q_1^l \oplus q_2^l \oplus \ldots \oplus q_n^l \oplus x_1^l \oplus x_2^l \oplus \ldots \oplus x_n^l$. Although $P_0$ also knows the values of $k_l$ and $q_1^l, q_2^l, \ldots, q_n^l$, this information is still helpless for her to obtain the value of $x_n^l$. Thus, $P_0$ cannot extract the secret of $P_n$.

It can be concluded now that $P_0$ has no knowledge about the secret of each party. Note that $P_0$ knows the comparison result of equality.

## 5 Discussion and conclusion

We compare the proposed MQPC protocol with the MQPC protocols of Refs.[36-43] after ignoring the security check processes. Table 1 summarizes the performance of these protocols. Note that there are two MQPC protocols in Ref.[38], which are represented by Ref.[38]-A and Ref.[38]-B here, respectively. Likewise, there are two MQPC protocols in Ref.[43], which are represented by Ref.[43] with two TPs and Ref.[43] with one TP here, respectively.

In summary, in this paper, we propose a novel MQPC protocol for equality comparison by using $n$-level single-particle states, where the encoded particles are transmitted in a circular way. Here, $n$ parties employ the qudit shifting operation to encode their private secrets and can compare the equality of their private secrets within one time execution of protocol. We validate in detail that the proposed MQPC protocol can overcome both the outside attack and the participant attack. Specially, each party has no access to other parties' secrets; and TP has no chance to know each party's secret either.

Table 1 Comparison between the proposed MQPC protocol and the MQPC protocols of Refs.[36-42]

| Quantum state | Quantum measurement | Quantum measurement | Number of | Quantum technology | Type of TP | Comparison of size | Pre-shared | Number of times to |

| | | for TP | for parties | parties | used | relation | | QKD key | compare $n$ parties |
|---|---|---|---|---|---|---|---|---|---|
| The protocol of Ref.[36] | $n$-particle GHZ class state | No | single-particle measurement | $n$ | The entanglement correlation among different particles from one quantum entangled state | The first kind of semi-honest TP | No | No | 1 |
| The protocol of Ref.[37] | $d$-level $n$-particle entangled state | $d$-level single-particle measurement | No | $n$ | Quantum fourier transform and unitary operation | The first kind of semi-honest TP | No | No | 1 |
| The protocol of Ref.[38]-A | $d$-level $n$-particle entangled state and $d$-level two-particle entangled state | $d$-level single-particle measurement | $d$-level single-particle measurement | $n$ | Quantum fourier transform | The second semi-honest model of TP | No | No | 1 |
| The protocol of Ref.[38]-B | $d$-level two-particle entangled state | $d$-level two-particle collective measurement | No | $n$ | Unitary operation | The second semi-honest model of TP | No | Yes | 1 |
| The protocol of Ref.[39] | $d$-level $n+1$-particle cat state and $d$-level two-particle Bell state | $d$-level $n+1$-particle cat state measurement | $d$-level two-particle Bell state measurement | $n$ | Quantum entanglement swapping and unitary operation | The second semi-honest model of TP | No | No | 1 |
| The protocol of Ref.[40] | $d$-level $l$-particle entangled state | No | $d$-level single-particle measurement | $n$ | The entanglement correlation among different particles from one quantum entangled state | The second semi-honest model of TP | Yes | Yes | 1 |
| The protocol of Ref.[41] | $d$-level $n$-particle GHZ state and $l$-level $n$-particle GHZ state | $d$-level single-particle measurement | $d$-level single-particle measurement | $n$ | Unitary operation | The second semi-honest model of TP | Yes | No | 1 |
| The protocol of Ref.[42] | $n$-particle GHZ state | No | single-particle measurement | $n$ | The entanglement correlation among different particles from one quantum entangled state | The second semi-honest model of TP | No | No | 1 |
| The protocol of Ref.[43] with two TPs | $d$-level single-particle state | $d$-level single-particle measurement | No | $n$ | Unitary operation | The second semi-honest model of TP | Yes | Yes | 1 |
| The protocol of Ref.[43] with one TP | $d$-level single-particle state | $d$-level single-particle measurement | No | $n$ | Unitary operation | The second semi-honest model of TP | Yes | Yes | 1 |
| The proposed MQPC protocol | $n$-level single-particle state | $n$-level single-particle measurement | No | $n$ | Unitary operation | The second semi-honest model of TP | No | Yes | 1 |

## Acknowledgments

Funding by the Natural Science Foundation of Zhejiang Province (Grant No.LY18F020007) is gratefully acknowledged.